\documentclass[preprint]{elsarticle}
\usepackage{amssymb}
\usepackage{graphics}
\usepackage{amsmath}
\usepackage{amsfonts}
\usepackage{bm}
\usepackage[]{latexsym}
\usepackage{epsfig}
\usepackage{float}
\usepackage{dsfont}

\newcommand{\be}{\begin{equation}}
\newcommand{\ee}{\end{equation}}
\newcommand{\bea}{\begin{eqnarray}}
\newcommand{\eea}{\end{eqnarray}}
\newcommand{\brr}{\begin{array}}
\newcommand{\err}{\end{array}}
\newcommand{\bit}{\begin{itemize}}\newcommand{\eit}{\end{itemize}}
\newcommand{\ben}{\begin{enumerate}}\newcommand{\een}{\end{enumerate}}

\newcommand{\ba}{\begin{array}}
\newcommand{\ea}{\end{array}}

\def\lf{\left}

\def\pa{\partial}

\def\ri{\right}

\def\al{\alpha}

\def\la{\lambda}
\def\si{\sigma}

\def\1{{_{1}}}\def\2{{_{2}}}

\begin{document}
\title{Comment on ``Neutrino decoherence in presence of strong gravitational fields'' }

\author[a,b]{L.~Petruzziello}
\ead{lupetruzziello@unisa.it}

\address[a]{INFN, Gruppo collegato di Salerno, Italy}
\address[b]{Dipartimento di Ingegneria, Universit\`a di Salerno, Via Giovanni Paolo II, 132 84084 Fisciano, Italy}

  \def\be{\begin{equation}}
\def\ee{\end{equation}}
\def\al{\alpha}
\def\bea{\begin{eqnarray}}
\def\eea{\end{eqnarray}}

  \def\be{\begin{equation}}
\def\ee{\end{equation}}
\def\al{\alpha}
\def\bea{\begin{eqnarray}}
\def\eea{\end{eqnarray}}

\begin{abstract}
The present work can be regarded as a corollary to the paper titled ``Neutrino decoherence in presence of strong gravitational fields''. Specifically, we investigate the gravitationally-induced decoherence on neutrino oscillation from the point of view of a locally inertial observer at rest with neutrino propagation. In so doing, we observe that in the aforementioned reference frame the effects of gravity can become manifest in a non-trivial manner with respect to an observer placed at infinity. 
\end{abstract}

 \vskip -1.0 truecm
\maketitle

\section{Introduction}

In a recent article appeared in literature~\cite{volpe}, the authors study the gravitational influence on neutrino decoherence in the wave packet (WP) approach. To this aim, they perform a thorough analysis of WP spreading due to the different ``detection'' point in curved spacetime for each mass eigenstate by resorting to the density matrix formalism. In particular, the attention is devoted to the radial propagation for the \emph{exact} Schwarzschild solution, so as to account also for the strong gravity regime, in which the gravitational effects on neutrino decoherence are expected to be prominent. As a matter of fact, the numerical esteem of the discrepancy between the coherence proper time in Schwarzschild spacetime and the coherence length in flat spacetime as a function of the source mass $M$ clearly indicates a significant shift even for $M\simeq1.4\,M_\odot$.

However, the only drawback in the above picture is represented by the employment of the unphysical quantity $r_{PD}=r_D-r_P$ in the expression of the density matrix, which becomes meaningful as long as $M\to0$. Additionally, the presence of the energy at infinity does not allow for a straightforward interpretation, which instead can be achieved by relying on a description in terms of a local inertial observer at rest with neutrino propagation. Along this line, examples that probe the worth of the previous prescription can be found in Refs.~\cite{our}, where neutrino oscillations are investigated in the local inertial frame. In the present paper, we want to perform a reasoning analogous to the one elucidated in Refs.~\cite{our} with the purpose of shedding light on the most important aspects of neutrino decoherence ascribable to gravity. Thus, the upcoming discussion can be regarded as a corollary to the results obtained in Ref.~\cite{volpe} rather than a criticism. To keep our considerations simple and at the same time explicit, in what follows we deal with the weak-field approximation\footnote{Such a choice does not spoil the generality of our considerations, in that the qualitative behavior of our results is expected to be correct also beyond the current approximation.}, since in this limit the outcome can be exhibited analytically.

As in Ref.~\cite{volpe}, throughout the work we use the units $\hbar=G=c=1$ and the mostly-positive signature convention for the metric tensor $(-,+,+,+)$. 

\section{Neutrino WP decoherence in Schwarzschild spacetime}

Let us introduce the line element for the weak-field approximation of the Schwarzschild solution
\be\label{metric}
ds^2=-\lf(1+2\phi\ri)dt^2+\lf(1-2\phi\ri)\lf(dr^2+r^2d\theta^2+r^2\sin^2\theta d\varphi^2\ri)\,,
\ee
where $\phi=-M/r$ is to be intended as a small parameter, and thus we restrict our attention to the linear contributions $\mathcal{O}(\phi)$. In order to study a locally inertial reference frame, we need to introduce vierbein fields, defined as~\cite{gravitation}
\be\label{tetrad0}
g_{\mu\nu}e^\mu{}_{\hat{a}}e^\nu{}_{\hat{b}}=\eta_{\hat{a}\hat{b}}\,,
\ee
where $\eta_{\hat{a}\hat{b}}$ is the Minkowski metric. In the above equation and henceforth, with the Greek indexes we denote manifold coordinate indexes, whereas the Latin and hatted ones are related to the ``Lorentzian'' vierbein labels. In the case under examination, the non-vanishing tetrads are given by
\be\label{tetrad}
e^t{}_{\hat{0}}=1-\phi\,, \qquad e^i{}_{\hat{j}}=\lf(1+\phi\ri)\delta^i_j\,.
\ee  
By virtue of the above formula, we immediately observe that the local energy $E_\ell$ can be written in terms of the one at infinity $E$ as~\cite{our,cardall}
\be\label{locale}
E_\ell=e^t{}_{\hat{0}}E=\lf(1-\phi\ri)E\,.
\ee
At this stage, we have to compute the dynamical covariant phase acquired by the $j$-th mass eigenstate during the propagation from the starting point $P$ to the detection point $D_j$, which is 
\be\label{cp}
\Phi_j(P,D_j)=\int_P^{D_j}\hspace{-1mm}p_\mu^{(j)}dx^\mu\,.
\ee
By keeping the analysis up to $\mathcal{O}(\phi)$ and following the assumptions of Ref.~\cite{volpe} (i.e. radial motion on the equatorial plane $\theta=\pi/2$), we find that
\be\label{cp2}
\Phi_j(P,D,\vec{p})=-E_j\lf(\vec{p}\ri)\lf(t_{PD}-b_{PD}\ri)-\frac{m_j^2}{2E_j(\vec{p})}r_{PD}\,,
\ee
where $t_{PD}=t_D-t_P$, $r_{PD}=r_D-r_P$ and
\be\label{b}
b_{PD}=r_{PD}-2\int_{r_P}^{r_D}\hspace{-1mm}\phi\,dr\,.
\ee
Next, we can compute the phase shift that enters the components of the density matrix $\rho_{jk}$ along with the elements of the mixing matrix and the (non-covariant) Gaussian WPs. Specifically, we have
\be\label{ro}
\rho_{jk}(P,D)=N^\al_{jk}\int\frac{d^3p}{(2\pi)^3}\int\frac{d^3q}{(2\pi)^3}e^{-i\Phi_{jk}}e^{-\frac{(\vec{p}-\vec{p}_j)^2}{4\si_p^2}}e^{-\frac{(\vec{q}-\vec{q}_k)^2}{4\si_p^2}}\,,
\ee
where $N_{jk}^\al$ is a constant containing the mixing matrix factors~\cite{volpe} and
\be\label{cp3}
\Phi_{jk}=E_{jk}\lf(t_{PD}-b_{PD}\ri)+\lf(\frac{m_j^2}{2E_j}-\frac{m_k^2}{2E_k}\ri)r_{PD}+\vec{v}_j\cdot(\vec{p}-\vec{p}_j)(t_{PD}-\la_j)-\vec{v}_k\cdot(\vec{q}-\vec{q}_k)(t_{PD}-\la_k)\,.
\ee
In the previous formulas, we have implicitly assumed to deal with a Gaussian WP centered around $\vec{p}_j$ with width $\si_p$ and~\cite{volpe,akl}
\be\label{req}
E_j=E_j(\vec{p}_j)\,, \qquad v_j=\lf.\frac{\pa E_j}{\pa p}\ri|_{p=p_j}\hspace{-2mm}\simeq1-\frac{m_j^2}{2E^2}\,, \qquad \la_j=\frac{m_j^2}{2E_j^2}r_{PD}+b_{PD}\,,
\ee
with a similar set of identities valid for the $k$-th mass eigenstate, whilst $E$ is the average energy between the two different mass eigenstates.

To approach the final outcome, we must integrate the quantity in Eq.~\eqref{ro} over time. Apart from an irrelevant constant and the usual dynamical phase (which can be found in Refs.~\cite{our,cardall}), we stress that the contribution responsible for the damping takes the form
\be\label{damp}
\rho_{jk}^{damp}(r_P,r_D)=\mathrm{exp}\lf[-\frac{\Delta m_{kj}^4}{32\,\si_x^2\,E^4}\,r_{PD}^2\ri]\,,
\ee 
with $\si_x=1/2\si_p$ and $\Delta m_{kj}^2=m_k^2-m_j^2$.

If we want the physical length $L_p$ to appear in Eq.~\eqref{damp}, we have to substitute $r_{PD}^2$ in terms of 
\be\label{pd}
L_p=\int_{r_P}^{r_D}\hspace{-1mm}\sqrt{g_{rr}}\,dr=\int_{r_P}^{r_D}\hspace{-1mm}(1-\phi)dr=r_{PD}\lf(1+\frac{M}{r_{PD}}\ln\frac{r_D}{r_P}\ri)\,,
\ee
but this turns out to be a trivial task in the approximation we are currently dealing with, namely
\be\label{pd2}
r_{PD}^2\simeq L_p^2\lf(1-2\,\frac{M}{L_p}\ln\frac{r_D}{r_P}\ri)\,.
\ee
Consequently, by suitably combining Eqs.~\eqref{locale}, \eqref{damp} and \eqref{pd2}, we can evaluate the damping term of the density matrix as seen by a locally inertial observer and written as a function of the physical length, that is
\be\label{damp2}
\rho_{jk}^{damp}=\mathrm{exp}\lf[-\frac{\Delta m_{kj}^4\,L_p^2}{32\,\si_x^2\,E_\ell^4}\lf(1-2\,\frac{M}{L_p}\ln\frac{r_D}{r_P}-4\,\frac{M}{r_D}\ri)\ri]\,.
\ee
The term $M/r_D$ only represents a constant factor that depends on the detection point. On the other hand, the interesting feature is the emergence of an extra term in addition to the usual (i.e. ``flat'') one that scales as $L_p$ rather than $L_p^2$ which can be unambiguously attributed to the gravitational field generated by the source mass $M$. Furthermore, we note that the direction of neutrino propagation is also significant, in that it can either enhance or diminish the damping process. In particular, it is possible to observe that if the mixed particle departs from $M$ (i.e. $r_D>r_P$), then the second term of Eq.~\eqref{damp2} is positive, thereby smoothing the overall damping. This can be simply explained by the fact that the propagating neutrino travels through a region of space where the gravitational influence becomes increasingly weaker. Conversely, for the opposite scenario (i.e. $r_D<r_P$) decoherence is improved due to the negative sign of the correction coming from gravity.  

As a final remark, we point out that, because of Eq.~\eqref{damp2}, the coherence length is also modified. Indeed, the density matrix is now suppressed by a factor $e^{-1}$ after a proper distance
\be\label{cl}
L^{coh}_{PD}=\frac{4\sqrt{2}\,E_\ell^2}{\lf|\Delta m_{kj}^2\ri|}\lf(1+2\,\frac{M}{r_D}\ri)\si_x+M\ln\frac{r_D}{r_P}\,.
\ee

\section{Conclusions}
In this corollary to Ref.~\cite{volpe}, we have explored the implications of gravitational decoherence on neutrino physics from the point of view of a locally inertial observer. We have shown the effects of an external gravitational field on the neutrino WP in such a reference frame, and we have highlighted the main features of the damping term arising in the density matrix. 

It is worth observing that the investigation carried out in the locally inertial reference frame is useful not only for a better understanding of the analyzed problem, but also for experimental and phenomenological purposes. For instance, the feasibility of gravitational tests involving neutrino interferometry~\cite{giunti} necessarily entails the employment of physical quantities that can only become manifest from the viewpoint of a local inertial observer. From a theoretical perspective, the interplay between neutrino physics and General Covariance gives rise to challenging debates centered around the importance of fundamental principles in Physics, as it can be found out in the recent literature~\cite{acc1,acc2,acc3}. In conjunction with topical issues borrowed from quantum information such as entanglement and decoherence\footnote{Interestingly, we stress that such concepts have already been exploited in the context of mixed particles~\cite{mix,mix2}.}, the above arguments may potentially open a window towards novel insights on the r\^{o}le that gravity plays in quantum systems and their practical applications. 

We emphasize one more time that the result~\eqref{damp2} can be analytically obtained only in the weak-field limit, otherwise the computation should have been performed numerically, thus becoming more tangled. Despite this, it is not difficult to envisage a qualitative behavior analogous to the one exhibited here also in the strong gravity regime.

\section*{References}

\end{document}